\RequirePackage{rotating}
\documentclass[twocolumn, tighten]{aastex631}
\usepackage{amsmath}
\usepackage[caption=false]{subfig}
\usepackage{xfrac}
\usepackage{CJK}

\shorttitle{Spiral arms in Haro 6-13}
\shortauthors{Huang et al.}

\begin{document}
\begin{CJK*}{UTF8}{gbsn} 
\title{Grand Design Spiral Arms in the Compact, Embedded Protoplanetary Disk of Haro 6-13}

\correspondingauthor{Jane Huang}
\email{jane.huang@columbia.edu}
\author[0000-0001-6947-6072]{Jane Huang}
\affiliation{Department of Astronomy, Columbia University, 538 W. 120th Street, Pupin Hall, New York, NY 10027, USA}
\author[0000-0001-8877-4497]{Masataka Aizawa}\affiliation{College of Science, Ibaraki University, 2-1-1 Bunkyo, Mito, Ibaraki 310-8512, Japan}
\author[0000-0001-7258-770X]{Jaehan Bae}
\affiliation{Department of Astronomy, University of Florida, Gainesville, FL 32611, United States of America}
\author[0000-0003-2253-2270]{Sean M. Andrews} \affiliation{Center for Astrophysics \textbar\ Harvard \& Smithsonian, 60 Garden St., Cambridge, MA 02138, USA}
\author[0000-0002-7695-7605]{Myriam Benisty}
\affiliation{Max Planck Institute for Astronomy (MPIA), Königstuhl 17, 69117 Heidelberg, Germany}
\author[0000-0003-4179-6394]{Edwin A. Bergin}
\affiliation{Department of Astronomy, University of Michigan, 323 West Hall, 1085 S. University Avenue, Ann Arbor, MI 48109, United States of America}
\author[0000-0003-4689-2684]{Stefano Facchini}
\affiliation{Dipartimento di Fisica, Universit\`a degli Studi di Milano, Via Celoria 16, 20133 Milano, Italy}
\author[0000-0002-4438-1971]{Christian Ginski}\affiliation{School of Natural Sciences, Center for Astronomy, University of 766 Galway, Galway, H91 CF50, Ireland}
\author[0000-0002-6338-3577
]{Michael K\"uffmeier}
\affiliation{Centre for Star and Planet Formation, Niels Bohr Institute, University of Copenhagen, \O ster Voldgade 5–7, 1350 Copenhagen, Denmark }

\begin{abstract}
Millimeter continuum spiral arms have so far only been detected in a handful of protoplanetary disks, and thus we have a limited understanding of the circumstances in which they can form. In particular, substructures in small disks ($R\lessapprox 50$ au) have not been well-characterized in comparison with large disks. We present ALMA 1.3 mm continuum observations of the disk around the T Tauri star Haro 6-13 at a resolution of $\sim0\farcs04$ ($\sim5$ au). A pair of low-contrast spiral arms are detected at disk radii from $\sim10-35$ au. They can be approximated as Archimedean spirals with pitch angles ranging from $\sim10-30^\circ$. The low value of the disk-averaged spectral index between 1.3 and 3 mm ($\alpha=2.1$) and the high brightness temperatures suggest that the millimeter continuum is likely optically thick and thus may hide sufficient mass for the disk to become gravitationally unstable and form spiral arms. CO observations have shown that Haro 6-13 is surrounded by an envelope, raising the possibility that infall is facilitating spiral arm formation.  
\end{abstract}


\section{Introduction} \label{sec:intro}

In the past decade, high resolution millimeter continuum observations have unveiled complex dust emission morphologies in numerous  protoplanetary disks \citep[e.g.,][]{2018ApJ...869L..41A, 2018ApJ...869...17L,2021MNRAS.501.2934C,2024ApJ...976..132H}. Axisymmetric gaps and rings are the most frequently detected type of substructure, and are commonly (though not universally) hypothesized to be due to planet-disk interactions \citep[e.g.,][]{2018ApJ...869L..42H, 2023ASPC..534..423B}. Millimeter continuum spiral arm detections are comparatively rare; they have been observed in a handful of young, embedded Class 0/I disks \citep[e.g.,][]{2014ApJ...796....1T, 2016Natur.538..483T, 2020ApJ...898...10T, 2020NatAs...4..142L, 2023ApJ...954..190X} and older Class II disks \citep[e.g.,][]{2016Sci...353.1519P, 2018ApJ...860..124D, 2018ApJ...869L..43H, 2018ApJ...869L..44K,  2020MNRAS.491.1335R, 2021ApJ...916...51A}. 

In addition to millimeter continuum, spiral arms have also been observed in disks in scattered light and molecular emission \citep[e.g.,][]{2004ApJ...605L..53F, 2012ApJ...748L..22M, 2014ApJ...785L..12C, 2017ApJ...840...32T, 2023ASPC..534..605B,2023ASPC..534..423B}. Scattered light and molecular emission trace surface and intermediate-height layers in the disk, whereas millimeter continuum emission originates from the dense midplane, where planet formation can take place. Spiral arms observed in one tracer do not necessarily have counterparts in others, suggesting that multiple spiral arm formation processes may be operating in disks, with varying influence in different vertical layers \citep[e.g.,][]{2018AA...620A..94G, 2018ApJ...869L..43H, 2023ASPC..534..423B}. 

Diverse origins have been proposed for spiral arms associated with protostellar/protoplanetary disks. The dynamical influence of binary interactions or stellar flybys may be responsible for some observed spiral structures \citep[e.g.,][]{2014ApJ...796....1T, 2016ApJ...816L..12D, 2018ApJ...869L..44K, 2020ApJ...896..132Z, 2020MNRAS.491.1335R}, but most disks with reported detections of spiral arms have no known companions \citep[e.g.,][]{2023ASPC..534..423B}. Along similar lines, spiral arms in Class II disks are commonly hypothesized to be associated with a planetary perturber \citep[e.g.,][]{2012ApJ...748L..22M, 2017ApJ...839L..24M, 2022ApJ...934L..11V}, with directly imaged (candidate) companions reported in the HD 100546, AB Aur, HD 169142, and MWC 758 systems \citep[e.g.,][]{2013ApJ...766L...1Q, 2022NatAs...6..751C, 2023MNRAS.522L..51H, 2023NatAs...7.1208W}. Some systems have extended non-Keplerian spiral structures that appear to trace material infalling onto their disks \citep[e.g.,][]{2017AA...603L...3A, 2021ApJ...908L..25G,2022ApJ...930...91D}. Several disks with spiral arms have been inferred to be gravitationally unstable either based on disk mass estimates and/or non-Keplerian kinematics \citep[e.g.,][]{2016Natur.538..483T, 2020NatAs...4..142L, 2021ApJ...914...88P, 2023MNRAS.518.4481L, 2023ApJ...954..190X, 2024Natur.633...58S}. In the Class 0 system L1448 IRS3B, gravitational instability (GI) has resulted in disk fragmentation to form a multiple star system. Meanwhile, for the detections in Class I and II disks, GI is of interest because it has been proposed as a pathway for forming planets either via fragmentation \citep[e.g.,][]{1998ApJ...503..923B,2004ApJ...609.1045M} or by enhancing the formation rate of large solids by trapping dust grains in spiral arms \citep[e.g.,][]{2023MNRAS.519.2017L, 2024MNRAS.528.2490R}. Identifying where and how spiral arms form is thus key for understanding what processes shape the structure and evolution of disks and set the conditions for planet formation. 

Haro 6-13 (ICRS coordinates RA=04:32:15.417, DEC= +24:28:59.58) is a K5.5 T Tauri star located $127.7\substack{+1.7\\-1.6}$ pc away in the Taurus Molecular Cloud complex \citep{2014ApJ...786...97H, 2021AA...649A...1G, 2021AJ....161..147B}. Classifications of this source based on its spectral energy distribution have varied. It was designated as a Class I/II young stellar object (YSO) by \citet{1995ApJS..101..117K}, a flat spectrum YSO by \citet{2010ApJS..186..259R}, and a Class II YSO by \citet{2010ApJS..186..111L}, who did not use the flat spectrum category. In any case, ALMA CO observations have shown that Haro 6-13 has an envelope \citep{2021AA...645A.145G}. \citet{2019ApJ...882...49L} estimated an age of $2.6\substack{+2.1\\-1.1}$ Myr based on comparisons of its stellar luminosity and effective temperature to evolutionary models. While age estimates of individual pre-main sequence stars are highly uncertain, Haro 6-13's estimated age  suggests that it could be an example of a system undergoing ``late infall,'' wherein a relatively evolved disk continues to accrete material from a remnant natal envelope or from encounters with nearby cloud material \citep[e.g.,][]{2019AA...628A..20D,2023EPJP..138..272K}. Models have predicted that infall may facilitate the formation of different kinds of disk substructures, including gaps and rings, spiral arms, and vortices \citep[e.g.,][]{2010ApJ...713.1143Z, 2015ApJ...805...15B, 2015AA...582L...9L, 2022ApJ...928...92K}. Recent scattered light and molecular line observations of other systems provide some support for a connection between late infall and spiral formation in the disk upper layers \citep[e.g.,][]{2021ApJ...908L..25G, 2021ApJS..257...19H, 2022AA...658A..63M}, but the effect on the disk midplane has been less explored. 

\citet{2019ApJ...882...49L} previously observed the millimeter continuum of Haro 6-13 at a resolution of $0\farcs1$ (13 au) as part of a survey of Class II disks in the Taurus star-forming region. No clear substructure was identified in their image, and thus they characterized Haro 6-13 as a ``smooth,'' compact disk with an estimated effective radius of $R_\mathrm{95}=34$ au, where $R_\mathrm{95}$ is the radius enclosing 95\% of the flux \citep{2019ApJ...882...49L}. At the resolution of their survey, the smaller disks appear to be generally more substructure-poor in comparison with the larger disks, and no spiral detections were reported for any system. However, non-axisymmetric residuals become apparent after they subtracted an axisymmetric intensity profile from the Haro 6-13 continuum. Based on 1D parametric visibility modeling of the same data, \citet{2023ApJ...952..108Z} inferred that the millimeter continuum of Haro 6-13 has a gap at 16 au and ring at 20 au. Meanwhile, using super-resolution imaging of this data, \citet{2024PASJ...76..437Y} inferred an inflection point, but not a gap (i.e., a local minimum) at a radius of 14 au. To investigate the nature of the substructures in the Haro 6-13 disk, we used ALMA to obtain high angular resolution 1.3 mm continuum observations. The observations are described in Section \ref{sec:observations}. Models and analysis of the system are presented in Section \ref{sec:analysis}. We discuss the results in Section \ref{sec:discussion} and close with a summary in Section \ref{sec:summary}. 

\section{Observations and Data Reduction}
\label{sec:observations}
Long-baseline 1.3 mm continuum observations of Haro 6-13 were obtained as part of ALMA program 2022.1.01365.S (PI: J. Huang). The correlator was configured with four spectral windows (SPWs) centered at 224, 226, 240, and 242 GHz, each of which had 128 channels across a bandwidth of 2 GHz. Details of the observing setup are listed in Table \ref{tab:observations}. A calibrated measurement set was produced by ALMA staff using the \texttt{CASA 6.4.1} pipeline
\citep{2022PASP..134k4501C}

\begin{deluxetable*}{lccccc}
\tabletypesize{\scriptsize}
\tablecaption {Observational setup \label{tab:observations}}
\tablehead{\colhead{Program ID} &\colhead{Date}&\colhead{Baseline lengths} & \colhead{Number of antennas}&\colhead{Time on source} & \colhead{Calibrators}}
\startdata
2016.1.01042.S & 2016 November 28 & 15.1 m - 704.1 m & 47 & 1.2 min& J0510+1800, J0438+3004\\
2016.1.01042.S & 2017 August 02 &  54.5 m - 3.3 km & 44 & 3 min & J0510+1800, J0423-0120, J0426+2327\\
2022.1.01365.S & 2023 July 19 & 230.2 m - 15.2 km & 47 & 32 min & J0429+2724, J0433+2905, J0510+1800 
\enddata 
\end{deluxetable*}

Observations in a more compact antenna configuration were requested as part of the same program but ultimately were not taken. To provide coverage on shorter baselines, we retrieved 1.3 mm observations of Haro 6-13 from program 2016.1.01042.S (PI: C. Chandler) via the ALMA archive. Details of the observing setup are listed in Table \ref{tab:observations}. The correlator was configured with 8 SPWs. Two SPWs were centered at 232.4 and 217.0 GHz, respectively, each with bandwidths of 2 GHz and 128 channels. The other SPWs each had bandwidths of 468.75 MHz. They were centered at 231.2 GHz (120 channels), 230.7 GHz (960 channels, covering $^{12}$CO $J=2-1$), 218.8 GHz and 219.3 GHz (60 channels each), and 219.7 and 220.2 GHz (480 channels each, covering C$^{18}$O and $^{13}$CO $J=2-1$, respectively). The raw data from 2016 November 28 were calibrated with the \texttt{CASA 4.7.2} pipeline, while the data from 2017 August 02 were processed using manual calibration scripts from the archive. 

Further processing of the datasets was performed with \texttt{CASA} 6.5. We began by flagging the channels covered by CO isotopologue line emission in the short-baseline measurement sets from program 2016.1.01042.S and averaged the channels in each SPW to generate pseudo-continuum visibilities. Separate continuum images were generated for the two execution blocks (EBs) using \texttt{tclean}. Following the procedure described in \citet{2018ApJ...869L..41A}, we used the \texttt{imfit} task in \texttt{CASA} to fit a 2D Gaussian to each continuum image in order to locate the disk center, then used the \texttt{phaseshift} task to shift the disk center to the phase center and \texttt{fixplanets} to match the coordinates of the two execution blocks (EBs). We combined the two EBs and performed three rounds of phase self-calibration with intervals corresponding to the scan length, 60 s, and 15 s, and then one round of amplitude self-calibration with a scan-length solution interval. The full spectral resolution data were spatially aligned in the same manner as the continuum observations, and the self-calibration tables derived from the continuum data were applied. Continuum subtraction was performed in the $uv$ plane using \texttt{uvcontsub}. Multiscale CLEAN \citep{2008ISTSP...2..793C} was used to image $^{12}$CO at a velocity resolution of 0.65 km s$^{-1}$ and $^{13}$CO and C$^{18}$O at a resolution of 1.35 km s$^{-1}$ (reflecting the lower native resolution of the latter two lines). We used a robust value of 0.5 and applied a $uv$-taper of $0.3''$ to improve the signal-to-noise ratio of the fainter, larger-scale emission. Since the emission morphology is irregular, the \texttt{auto-multithresh} algorithm \citep{2020PASP..132b4505K} was used to create the CLEAN masks. After applying a primary beam correction to the image cubes, we produced integrated intensity maps of $^{12}$CO by summing up emission at LSRK velocities between 0.2 and 10.6 km s$^{-1}$ and of $^{13}$CO and C$^{18}$O by summing emission between 2.75 and 8.15 km s$^{-1}$, with negative pixels clipped to minimize artifacts from spatial filtering. The velocity ranges were chosen based on the detection of emission above the $3\sigma$ level. Intensity-weighted velocity maps were then produced using a 4$\sigma$ clip.

Next, we channel-averaged the SPWs from the long-baseline measurement set from 2022.1.01365.S and performed one round of phase self-calibration with a solution interval spanning the length of the observation. Then, the long-baseline observations were spatially aligned with the self-calibrated short-baseline continuum observations using the same procedure described above. The combined data were imaged with multiscale CLEAN and a robust value of 0.5. We performed two rounds of phase self-calibration on the combined data with scan-length and 30 s intervals, and then one round of amplitude self-calibration with a scan-length interval. Finally, a primary beam correction was applied. Image properties are listed in Table \ref{tab:imageproperties}.\footnote{The visibilities and images can be downloaded from Zenodo: doi:\href{https://zenodo.org/records/15002753}{10.5281/zenodo.15002753}}

\begin{deluxetable*}{lcccc}
\tabletypesize{\scriptsize}
\tablecaption {Image properties \label{tab:imageproperties}}
\tablehead{\colhead{Image} &\colhead{Frequency}&\colhead{Synthesized beam} & \colhead{rms\tablenotemark{a}}\\
& \colhead{(GHz)}&\colhead{(arcsec $\times$ arcsec (deg))}&\colhead{(mJy beam$^{-1}$)}}
\startdata
Continuum&229.496&$0\farcs037\times0\farcs022$ $ (19\fdg08)$&0.019\\
$^{12}$CO $J=2-1$&230.538&$0\farcs47\times0\farcs42$ $ (-31\fdg62)$&7.5\\
$^{13}$CO $J=2-1$&220.3986842&$0\farcs48\times0\farcs42$ $ (-32\fdg99)$&4.9\\
C$^{18}$O $J=2-1$ & 219.5603541 & $0\farcs47\times0\farcs42$ $ (-32\fdg20)$&4.0
\enddata 
\tablenotetext{a}{Measured from line-free channels for the image cubes ($\Delta v = 0.65$ km s$^{-1}$ for $^{12}$CO and 1.35 km s$^{-1}$ for $^{13}$CO and C$^{18}$O), and from an off-source region for the continuum emission.}
\end{deluxetable*}

\section{Analysis} \label{sec:analysis}
\subsection{Overview of continuum morphology}

\begin{figure*}
\begin{center}
\includegraphics{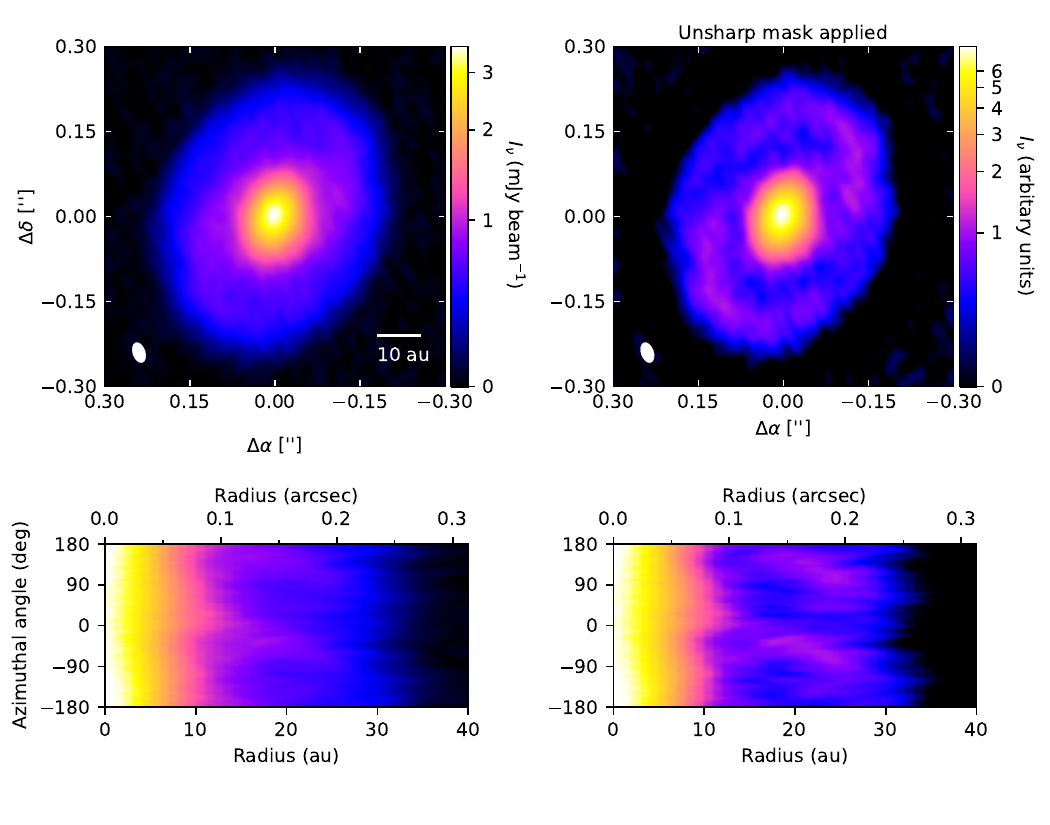}
\end{center}
\caption{Left: 1.3 mm continuum image of Haro 6-13 and the corresponding intensity map as a function of radius and azimuthal angle in disk coordinates. An arcsinh stretch is used for the color scale. The synthesized beam is shown in the lower left corner of the CLEAN image. Right: The CLEAN image with an unsharp mask applied and the corresponding intensity map in $R$, $\theta$ coordinates. \label{fig:continuumoverview}}
\end{figure*}

The 1.3 mm continuum image and corresponding intensity plot as a function of disk radius $R$ and azimuthal angle $\theta$ are shown in Figure \ref{fig:continuumoverview}. The derivation of the position angle (P.A.) and inclination ($i$) used for deprojection is described in Section \ref{sec:spiralproperties}. An angle of $\theta = 90^\circ$ corresponds to the side of the major axis that defines the position angle (east of north), and the angles increase in the clockwise direction. The continuum emission exhibits a faint pair of spiral arms. To enhance the spiral arm contrast, we applied unsharp masking, which subtracts a smoothed version of an image from the original image and then adds a scaled version of the difference image back to the original image. We used the implementation from \texttt{scikit-image}, specifying a scale factor of 2 and a Gaussian filter with a sigma parameter of $0\farcs06$. The resulting image and intensity plot in $R$ and $\theta$ are also shown in Figure \ref{fig:continuumoverview}.

\begin{figure}
\begin{center}
\includegraphics{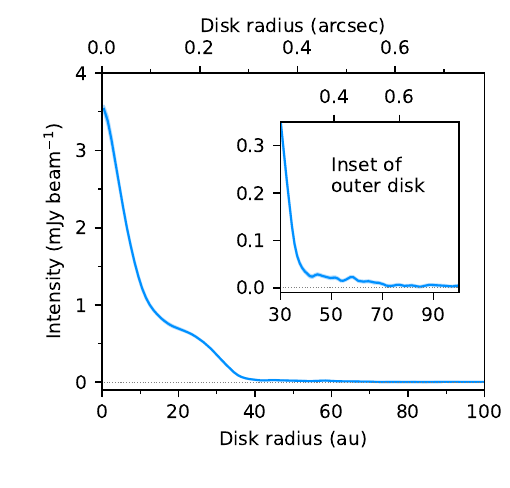}
\end{center}
\caption{Azimuthally averaged, deprojected radial profile of the 1.3 mm continuum image. The shaded ribbon shows the scatter in each radial bin divided by the square root of the number of synthesized beams spanning each bin. The inset shows the faint emission at larger radii.\label{fig:radialprofile}}
\end{figure}

Figure \ref{fig:radialprofile} shows the azimuthally averaged, deprojected radial intensity profile made from the original CLEAN image of Haro 6-13's continuum emission. The intensity profile drops steeply from the center of the disk until it flattens out at $R\approx15$ au, then drops steeply again beyond $R\approx25$ au. No clear gaps, rings, or cavities are visible, but it is possible that the slope change at $R\sim15$ au traces an unresolved gap. Beyond $R\approx40$ au, very faint emission is visible out to $R\approx70$ au. Without azimuthal averaging, this emission only appears as scattered $3\sigma$ peaks in the image, so it is not clear whether it originates from the Keplerian disk or from the envelope. Previous molecular line observations indicate that the Keplerian gas disk extends to a radius of 150 au \citep{2021AA...645A.145G}.

\subsection{Spiral arm properties\label{sec:spiralproperties}}
To isolate the spiral structures from the disk background, we estimated and subtracted an axisymmetric intensity profile using the publicly available code \texttt{protomidpy}\footnote{\url{https://github.com/2ndmk2/protomidpy}}, which implements the algorithm from \citet{2024MNRAS.532.1361A}. (We do not use the version of the image with unsharp masking for quantitative analysis because the process truncates the outer disk emission, which may bias spiral measurements). In brief, \texttt{protomidpy} assumes that the disk is axisymmetric and geometrically thin and models the 1D radial intensity profile $I(r)$ as a Fourier-Bessel series. The radial intensity profile model is regularized by a radial basis function kernel with hyperparameters $\alpha$ (which controls the weight of this prior) and $\gamma$ (which controls the spatial scale of the kernel). Model visibilities are calculated from the 1D radial intensity profile given the disk P.A., $i$, and offsets from the phase center ($\Delta x$ and $\Delta y$). \texttt{protomidpy} uses the affine invariant Markov Chain Monte Carlo (MCMC) code \texttt{emcee} \citep{2013PASP..125..306F} to sample the posteriors for P.A., $i$, $\Delta x$, $\Delta y$, $\alpha$, and $\gamma$ while simultaneously estimating $I(r)$. The non-parametric modeling framework of \texttt{protomidpy} shares some similarities with the \texttt{frank} code \citep{2020MNRAS.495.3209J}, but the latter requires the user to specify the disk orientation, offset, and regularization hyperparameters.  

Before modeling the visibilities with \texttt{protomidpy}, we used the \texttt{visread} package \citep{ian_czekala_2021_4432520}, which implements the method described in \citet{2023PASP..135f4503Z}, to correct the absolute scaling of the weights in our measurement sets on a per-SPW basis. For the various observations, the weights were ultimately downscaled by factors ranging from $\sim1$ (weights were essentially correct) to $\sim5$. 

\begin{figure*}
\begin{center}
\includegraphics{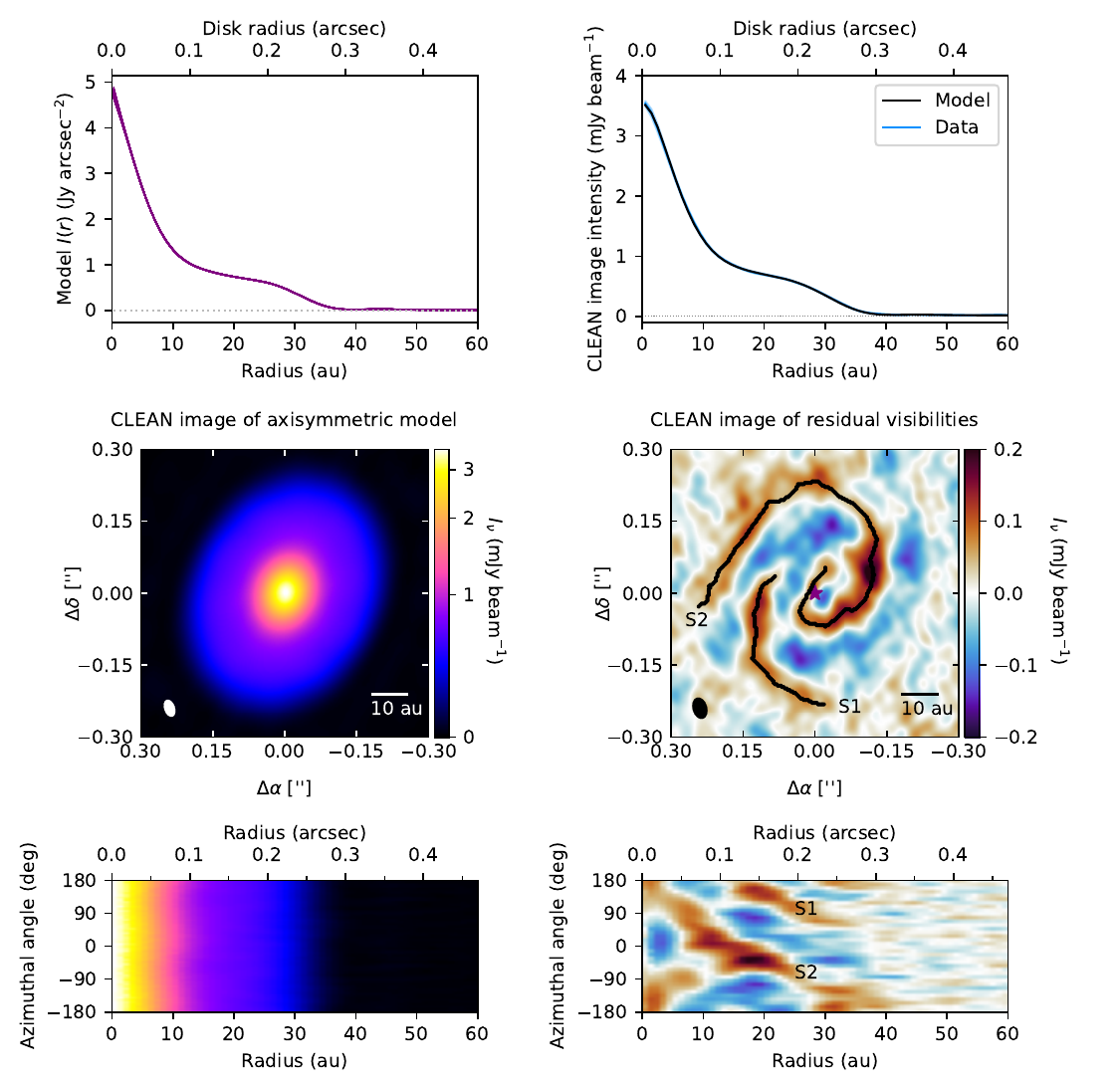}
\end{center}
\caption{Top left: 300 random $I(r)$ samples from \texttt{protomidpy} modeling. Center left: CLEAN image of axisymmetric \texttt{protomidpy} model. Bottom left: Corresponding intensity plot in $R$ and $\theta$ coordinates. Top right: Comparison of radial profiles calculated from CLEAN images of the observations and model. Center right: CLEAN image of the continuum residual visibilities created by subtracting the \textit{maximum a posteriori} \texttt{protomidpy} model from the observed visibilities. The purple star marks the disk center. The black dots mark the spiral arm positions identified by \texttt{filfinder}. Bottom right: Plot of the continuum residuals in $R$ and $\theta$ coordinates.  \label{fig:protomidpy}}
\end{figure*}

We set flat priors with bounds of $[0,1]$ for $\cos i$, $[0^\circ, 180^\circ]$ for P.A., $[-0\farcs1, 0\farcs1]$ for $\Delta x$ and $\Delta y$, $[0\farcs01, 0\farcs15]$ for $\gamma$, and $[-4,5]$ for $\log\alpha$. The MCMC modeling was performed using 24 walkers over 15000 steps. The length of the chains was chosen to ensure that they were at least $50\times$ longer than the integrated autocorrelation time.\footnote{\url{https://emcee.readthedocs.io/en/stable/tutorials/autocorr/}} After discarding the first 1000 steps for each walker as burn-in, we computed the medians 
and the 68\% confidence interval. The parameter values are given in Table \ref{tab:protomidpy}. Samples of the resulting $I(r)$ profiles inferred by \texttt{protomidpy} are plotted in Figure \ref{fig:protomidpy}. 

\begin{deluxetable}{cc}
\tablecaption {Inferred \texttt{protomidpy} parameter values \label{tab:protomidpy}}
\tablehead{\colhead{Parameter} &\colhead{Value}}
\startdata
$\Delta x$ (mas)&$3.61\pm0.06$\\
$\Delta y$ (mas) & $0.81\pm0.08$\\
P.A. (deg) & $154.09\pm0.14$ \\
$i$ (deg) & $39.47\pm0.09$\\
Hyperparameter $\gamma$ (arcsec) & $0.079\pm0.003$\\
Hyperparameter $\log \alpha$ & $0.5\pm0.2$ 
\enddata 
\end{deluxetable}

We then CLEANed the model visibilities generated by \texttt{protomidpy} from its \textit{maximum a posteriori} estimate of $I(r)$ in the same manner as the observations shown in Figure \ref{fig:continuumoverview}. Next, we CLEANed the residual visibilities created by subtracting the model visibilities from the observed visibilities, employing a $uv$ taper of $0\farcs02$ to improve the signal-to-noise ratio of the residual emission. The resulting synthesized beam is $0\farcs046\times0\farcs030$ $(15\fdg9)$. The CLEAN image of the axisymmetric \texttt{protomidpy} model, comparison of the radial profiles calculated from the CLEAN images of the model and observations, CLEAN image of the residual visibilities, and corresponding intensity plots in $R$ and $\theta$ coordinates are also presented in Figure \ref{fig:protomidpy}. The spiral that terminates on the southwestern side of the disk is labelled S1, while the one that terminates on the northeastern side of the disk is labelled S2. S2 appears to have both a greater inward and outer radial extent compared to S1. In addition, S2 appears to be connected to a bar-like structure at the center of the disk. However, tests on synthetic data from \citet{2024MNRAS.532.1361A} show that spiral structure can appear significantly distorted within a beam or so from the disk center, so higher resolution observations should be obtained to confirm whether a bar is genuinely present.

Next, we used \texttt{filfinder} \citep{2015MNRAS.452.3435K, 2016ascl.soft08009K} to extract the sky coordinates of the spiral arms from the \texttt{protomidpy} residual image. While \texttt{filfinder} was originally designed to identify filaments in maps of molecular clouds, \citet{2022ApJ...928....2I} demonstrated its utility in a protoplanetary disk context by identifying spiral structures in CO velocity residuals. After some experimentation, we set a smoothing size of 2 pixels (corresponding to $0\farcs006$), an intensity threshold of 0.03 mJy beam$^{-1}$, a threshold of 4 pixels (corresponding to $0\farcs012\sim1.5$ au) for the width of the adaptive thresholding mask, a size threshold of 300 au$^2$, and a minimum length of 3 synthesized beams. As shown in Figure \ref{fig:protomidpy}, \texttt{filfinder} identifies one structure corresponding to S1 and another structure that corresponds to S2 and the central bar-like residual.

\begin{figure}
\begin{center}
\includegraphics{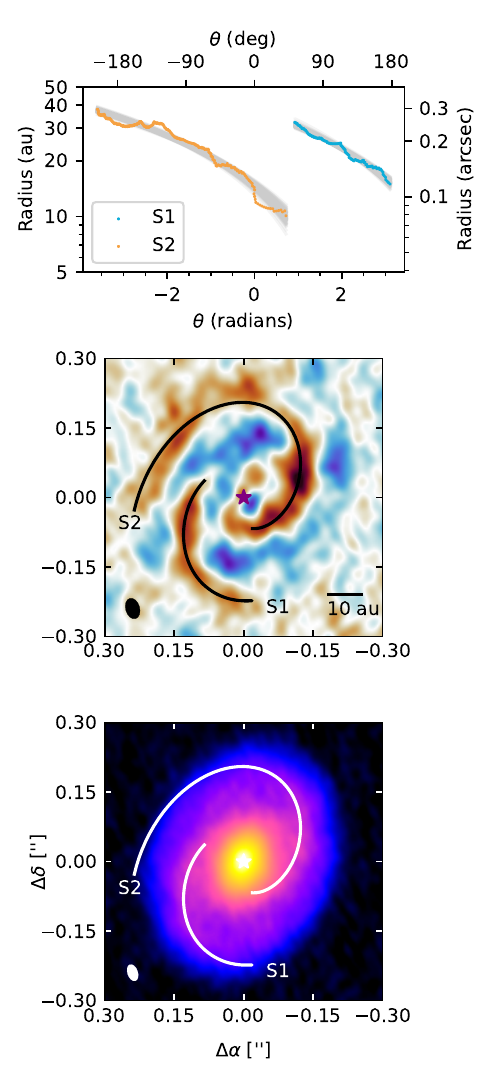}
\end{center}
\caption{Top: Spiral arm positions identified by \texttt{filfinder} (orange and blue dots) and Archimedean spirals generated with 100 random draws of $a$ and $b$ from the posteriors (gray curves). Middle: Archimedean spirals generated with the posterior median values of $a$ and $b$ for each arm, plotted on top of the $\texttt{protomidpy}$ residuals. Bottom: The same Archimedean spirals plotted over the original CLEAN image of the disk.  \label{fig:spiralfits}}
\end{figure}

The spiral coordinates identified by \texttt{filfinder} were then converted to disk polar coordinates $R$ and $\theta$, as shown in Figure \ref{fig:spiralfits}. The $\theta$ values are unwrapped to avoid jumps of $2\pi$, so for portions of the spirals, their values may differ by $2\pi$ from Figure \ref{fig:protomidpy}. For S2, the points identified by \texttt{filfinder} with $R<10$ au are not plotted because they are associated with the central bar-like residual. The pitch angle $\beta$ of a spiral arm can be calculated from
$\tan \beta = \left| \frac{d\ln R}{d\theta} \right|$. Thus, in a $\ln R$-$\theta$ plot, a spiral with constant pitch angle (a logarithmic spiral) will appear as a straight line, a spiral with a pitch angle that increases with $R$ will have a derivative with an absolute value that increases with $R$, while a spiral with a pitch angle that decreases with $R$ will have a derivative with an absolute value that decreases with $R$. S1 is only detected over a relatively small polar angle range (from $\theta\sim1-3$ radians), for which it is difficult to distinguish whether the pitch angle is varying \citep[e.g.,][]{2018ApJ...869L..43H, 2018ApJ...869L..44K}. However, S2 can be traced over a wider range, from $\theta\sim-3.5$ to 1 radian. S2 flattens out as $R$ increases, indicating that the pitch angle is decreasing with radius. We thus model S1 and S2 as Archimedean spirals, which take the form $R(\theta) = a-b\theta$. 

Adjacent pixels are not independent because the synthesized beam is oversampled. To account for correlated noise in our measurements of the spiral arm positions, we used the \texttt{george} package \citep{2015ITPAM..38..252A} to perform Gaussian process regression. The elements of the covariance matrix are modelled as the sum of a white noise component and a squared exponential kernel:

\begin{equation}
K_{ij} = \sigma^2 \delta_{ij}+A\exp\left[-\frac{(\theta_i-\theta_j)^2}{2C}\right]
\end{equation}
The squared exponential kernel specifies high covariance for datapoints that are close to one another and low covariance for datapoints that are far away from each other, which is qualitatively consistent with the behavior of pixels in an image with finite resolution. 

Each of the arms were fit independently for the spiral geometry parameters $\ln a$ and $\ln b$ and the hyperparameters $\ln(\sigma^2)$, $\ln A$, and $\ln C$. The fit is performed in log space because $a$, $b$, $\sigma^2$, $A$, and $C$ must all be positive. For $\ln a$, we set a flat prior with bounds of $[3,5]$ for S1 and $[1,4]$ for S2. For both arms, we set flat priors with bounds of $[1,3]$ for $\ln b$, $[-5, 5]$ for $\ln \sigma^2$, $[-4,4]$ for $\ln A$, and $[-10, 0]$ for $\ln C$. 

\begin{figure*}
\begin{center}
\includegraphics{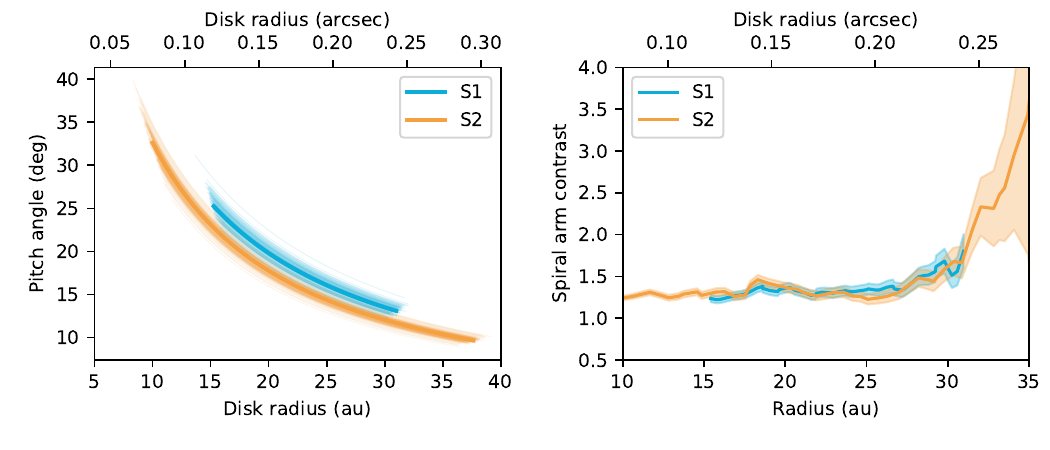}
\end{center}
\caption{Left: Pitch angles as a function of radius for the spiral arm models. The bold, darker curves correspond to the Archimedean spirals with $a$ and $b$ set to the posterior median values, while the lighter, thin curves show 100 random draws from the posterior for each arm. Right: Contrasts of each spiral arm. The shaded region shows the $1\sigma$ uncertainty. \label{fig:pitchangle}}
\end{figure*}

We used \texttt{emcee} to sample the posterior distributions. To model each arm, an ensemble of 30 walkers was evolved for 5,000 burn-in steps and then an additional 10,000 steps. Table \ref{tab:spiralparameters} lists the median and the $68\%$ confidence interval for each parameter. Figure \ref{fig:spiralfits} compares the data with the model Archimedean spirals, which appear to reasonably reproduce the observations. We then calculated the pitch angles as a function of radius from the spiral models (Figure \ref{fig:pitchangle}). For S1, the pitch angle ranges from $25^\circ$ at $R=15$ au to $13^\circ$ at $R=31$ au. For S2, the pitch angle ranges from  $33^\circ$ at $R=10$ au to $10^\circ$ at $R=38$ au. S1 appears to have a slightly larger pitch angle than S2 at the same radii, but the pitch angles of the two arms are consistent within the uncertainties.

\begin{deluxetable}{ccc}
\tabletypesize{\scriptsize}
\tablecaption {Spiral model parameters \label{tab:spiralparameters}}
\tablehead{\colhead{Parameter} &\colhead{S1}&\colhead{S2}}
\startdata
$a$ (au)& $37.6\pm0.7$&$14.6\pm0.5$\\
$b$ (au)& $7.2\pm0.3$&$6.4\pm0.2$\\
$\ln (\sigma^2/\text{au}^2)$ & $-3.8\pm0.2$&$-3.2\pm0.1$\\
$\ln (A/\text{au}^2)$&$-0.7\substack{+0.5\\-0.4}$&$0.8\substack{+0.3\\-0.2}$\\
$\ln (C)$&$-4.7\substack{+0.8\\-0.2}$ &$-4.9\pm0.1$ \\
\enddata 
\end{deluxetable}

We then measured the contrast of each spiral arm. The spiral arm intensities were extracted from the original CLEAN image of the disk (shown in Figure \ref{fig:continuumoverview}) along the coordinates of the best-fit Archimedean spiral arms. The ``background'' was estimated by measuring the minimum intensity of the disk in radial bins of 1 au and then interpolating at the radial locations of the pixels that fall along the model spiral arms. The contrast of each arm is then the spiral arm intensity divided by the ``background'' intensity. The uncertainty on the spiral arm and background intensity measurements was assumed to be the rms of the image. The contrasts are plotted in Figure \ref{fig:pitchangle}. Over the radii where both arms are detected, their contrasts are similar to one another and generally low ($<1.5$). The contrasts are nearly flat between radii of 10 and 30 au, but appear to rise for S2 beyond 30 au. This rise roughly coincides with the radius at which the radial intensity profile drops steeply. The increase in contrast is likely due to some combination of lower optical depths at larger radii as well as having a larger number of resolution elements along the azimuthal direction. 

Our spiral geometry measurements assume that the millimeter continuum disk is geometrically thin. While millimeter wavelength observations indicate that the large dust grains in Class II disks are often well-settled \citep[e.g.,][]{2022ApJ...930...11V, 2023MNRAS.524.3184P}, observations of younger Class 0 and I systems suggest that the millimeter wavelength emission sometimes comes from more elevated heights due to a lesser degree of settling \citep[e.g.,][]{2023ApJ...946...70V, 2023ApJ...951....8O}. Models have shown that infall can lead to enhanced turbulence levels in disks \citep[e.g.,][]{2022ApJ...928...92K, 2024ApJ...972L...9W}, which may inhibit vertical settling. \citet{2021ApJ...916...51A} demonstrated that for disks with elevated emission surfaces, subtracting a geometrically thin axisymmetric emission model leaves antisymmetric residuals on opposide sides of the major axis. This may contribute to the apparent difference in the extents of S2 and S1, as well as the bar-like residual near the disk center. 

\subsection{Evidence for high millimeter continuum optical depths}
From our observations, we measure a flux of 144.8 mJy at 229.5 GHz (1.3 mm) inside a circular region with a $1''$ radius. This value is consistent with the 1.3 mm flux from \citet{2019ApJ...882...49L}. Given the high SNR of our data, the dominant source of uncertainty is the 10\% systematic flux calibration uncertainty. Meanwhile,   \citet{2019ApJ...877L...2H} reported a flux of 24.96 mJy at 97.5 GHz (3 mm). The flux measurements at 229.5 and 97.5 GHz yield a disk-averaged spectral index of $\alpha = \ln\frac{I_{\nu_1}}{I_{\nu_2}}/\ln{\frac{\nu_1}{\nu_2}} = 2.1\pm0.2$. Because the 3 mm observations from \citet{2019ApJ...877L...2H} only marginally resolve the disk, we cannot meaningfully estimate a radially resolved spectral index profile. Using CARMA, \citet{2015ApJ...808..102K} measured disk fluxes of 111.2 mJy at 229 GHz and 28.7 mJy at 113 GHz. Again assuming a 10\% systematic flux calibration uncertainty, their measurements yield a disk-averaged $\alpha = 1.9\pm0.2$, which is consistent with our estimate from ALMA. 

Low spectral index values (near $\alpha\approx2$) can result from optically thick dust emission \citep[e.g.,][]{2012AA...540A...6R, 2019ApJ...877L..22L, 2019ApJ...877L..18Z}. Assuming a standard power-law grain size distribution and a DSHARP dust composition \citep{2018ApJ...869L..45B}, optically thin emission from large dust grains (maximum sizes of several millimeters or greater) can yield spectral index values as low as $\sim2.5$ \citep{2019ApJ...877L..18Z}. Our low spectral index value is compatible with the disk being optically thick, but given the uncertainties, optically thin emission from large dust grains is not ruled out based on spectral index alone. 

\begin{figure}
\begin{center}
\includegraphics{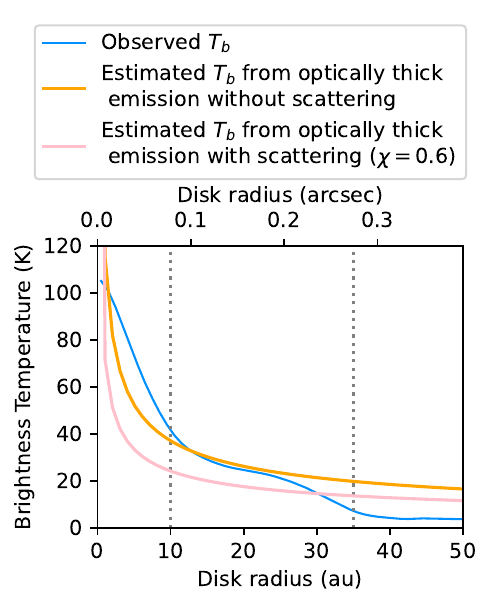}
\end{center}
\caption{A comparison of the observed brightness temperature profile of the Haro 6-13 disk to the expected brightness temperatures from optically thick emission in the absence of scattering (i.e., saturation at the dust temperature) or optically thick emission with high scattering ($\chi=0.6$). The gray vertical lines mark the approximate radial range over which spiral arms are detected. \label{fig:brightnesstemps}}
\end{figure}

However, the high brightness temperatures of our 1.3 mm continuum emission also point to high optical depths. In the absence of scattering, the brightness temperature should saturate at the dust temperature. With scattering, the observed intensity $I_\nu$ will saturate below blackbody values, with $\chi \equiv \frac{I_\nu}{B_\nu}$ as low as $\sim0.6$ for the DSHARP dust composition \citep{2019ApJ...877L..18Z}. We use the following analytic formula for a passively irradiated disk from \citet{2018ApJ...869L..46D} to estimate the dust temperature:
\begin{equation}\label{eq:Tdust}
T_d(R)  = \left( \frac{\varphi L_\ast}{8\pi R^2\sigma_\mathrm{SB}}\right)^{0.25}.
\end{equation}
As in \citet{2018ApJ...869L..46D}, we set the flaring angle $\varphi$ to 0.02. The luminosity of Haro 6-13 is $L_\ast = 0.79$ $L_\odot$ \citep{2019ApJ...882...49L}. Figure \ref{fig:brightnesstemps} compares the observed brightness temperatures (calculated from the CLEAN image using the full Planck function) to the brightness temperatures corresponding to our calculated dust temperatures for the cases of 1) optically thick emission with no scattering and 2) optically thick emission with $\chi=0.6$. Throughout most of the region in which spiral arms are detected, the observed brightness temperature is marginally below the estimated dust temperature and above the corresponding brightness temperature for the case with $\chi=0.6$ (high scattering). Interior to the spiral arms, the analytic dust temperature formula actually underestimates the brightness temperature, except in the inner few au, where beam dilution leads to an artificially low brightness temperature. In any case, while we only have a crude estimate of the dust temperature, it serves to demonstrate that the observed brightness temperatures are in the range of brightness temperatures expected when optical depths are high.  

\subsection{CO isotopologue emission}

\begin{figure*}
\begin{center}
\includegraphics{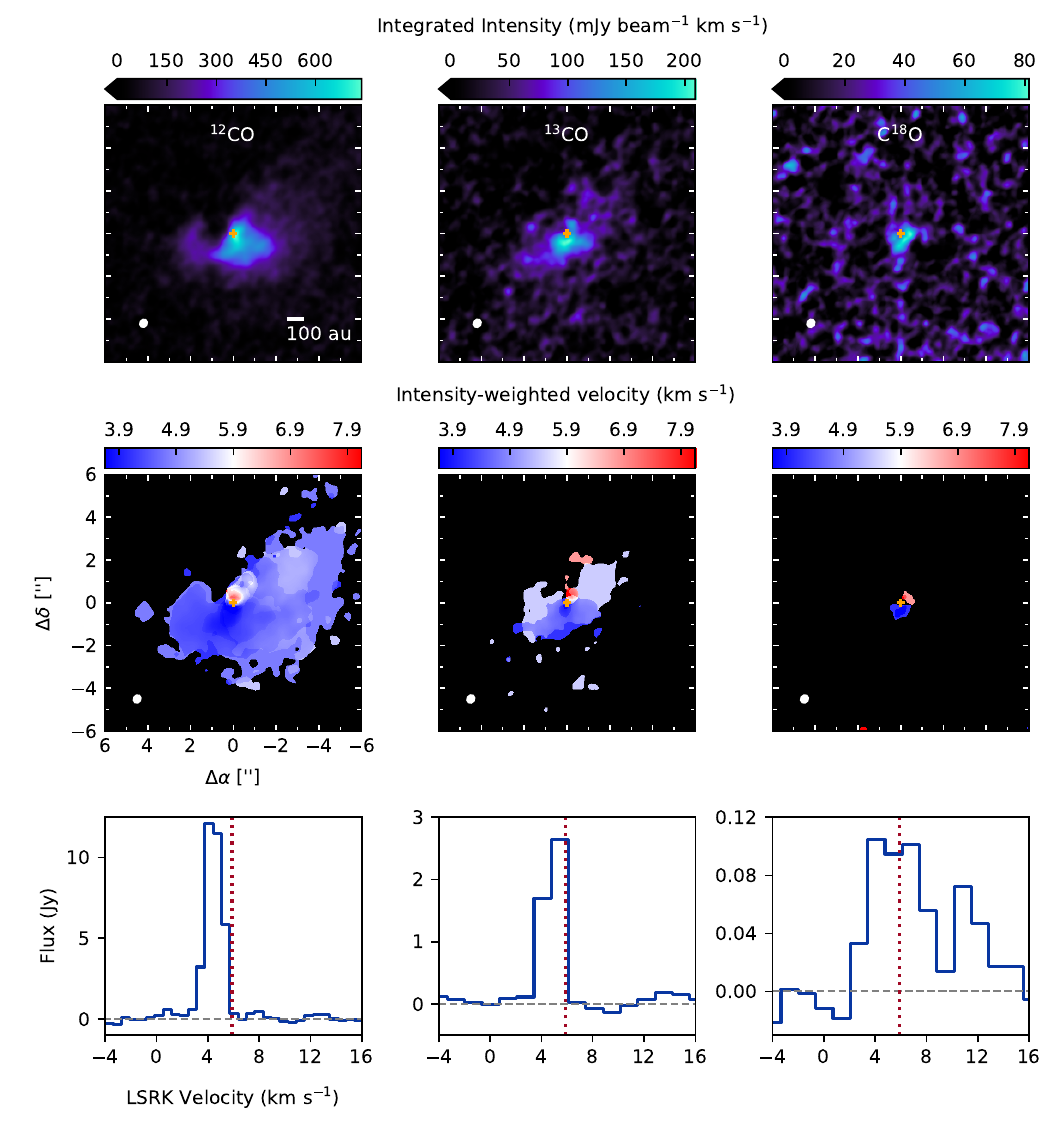}
\end{center}
\caption{Top row: Integrated intensity maps of $^{12}$CO, $^{13}$CO, and C$^{18}$O, respectively. The orange crosses mark the position of the center of the continuum disk. The synthesized beam is shown as a white ellipse in the lower left corner of each panel. Middle row: Intensity-weighted velocity maps of the three molecules. Bottom row: Spatially integrated spectra of the three molecules. The dotted red line denotes the systemic velocity of 5.9 km s$^{-1}$ \citep{2021AA...645A.145G}. \label{fig:COmaps}}
\end{figure*}

The integrated intensity maps, intensity-weighted velocity maps, and spectra of the CO isotopologues are shown in Figure \ref{fig:COmaps}. The $^{12}$CO and $^{13}$CO spectra were extracted from circular apertures centered on the disk with radii of $6''$ (770 au). Meanwhile, the spectrum of C$^{18}$O, which is detected over a smaller spatial extent,  was extracted with a circular aperture with a radius of $1\farcs5$. $^{12}$CO and $^{13}$CO are dominated by envelope emission extending to hundreds of au south and west of the star. The spectra show that the bulk of the observed emission is blueshifted with respect to the systemic velocity of 5.9 km s$^{-1}$ \citep{2021AA...645A.145G}, with much of the redshifted emission absorbed and/or resolved out, leading to the sharp drop in emission around the systemic velocity. The optically thinner C$^{18}$O emission is more symmetric about the systemic velocity, with a hint of a double-peaked Keplerian profile. Its intensity-weighted velocity map shows redshifted emission to the northwest of the star and blueshifted emission southeast of the star, consistent with the Keplerian disk rotation pattern observed in higher spectral resolution H$_2$CO and CS observations by \citet{2021AA...645A.145G}. However, as with $^{12}$CO and $^{13}$CO, the C$^{18}$O integrated intensity map also shows more extended blueshifted emission south of the star. Given the coarse spectral resolution of the data, it is not straightforward to distinguish disk and envelope contributions to the C$^{18}$O emission. However, the CS observations from \citet{2021AA...645A.145G} show a non-Keplerian blueshifted component south of the star, suggesting that envelope emission is also contributing to the C$^{18}$O emission asymmetry. The C$^{18}$O spectrum also shows a small spike in emission at $\sim11$ km s$^{-1}$ that is not observed in either $^{13}$CO or $^{12}$CO. Inspection of the image cube does not reveal any obvious coherent structure at this velocity, so this is likely to be a noise fluctuation.

\section{Discussion} \label{sec:discussion}
\subsection{Assessing possible explanations for Haro 6-13's spiral arms}
\subsubsection{Gravitational instability}
Spiral arms can form in disks when the Toomre $Q$ parameter \citep{1964ApJ...139.1217T} is $\lessapprox$2 \citep[e.g.,][]{2004MNRAS.351..630L, 2009MNRAS.393.1157C}. For a Keplerian disk, 
\begin{equation}
Q = \frac{c_s \Omega_K}{\pi G \Sigma},
\end{equation}
where $c_s =\sqrt{\frac{kT}{\mu m_H}}$ is the sound speed, $\mu=2.37$ is the mean molecular weight, $\Omega_K = \sqrt{\frac{GM_\ast}{R^3}}$ is the angular velocity at disk radius $R$, $G$ is the gravitational constant, and $\Sigma$ is the disk surface density. GI thus requires high disk surface densities, and is more readily maintained through the ongoing delivery of material via infall \citep[e.g.,][]{2010ApJ...713.1143Z,2011MNRAS.413..423H,2014ApJ...795...61B}. In addition, 3D hydrodynamical \citep{2016SciA....2E0875L} and magnetohydrodynamical zoom-in simulations \citep{2018MNRAS.475.2642K} demonstrated that infalling material can locally trigger GI in the disk. The presence of envelope material around Haro 6-13 therefore motivates the consideration of GI as the origin of its spiral arms. As noted in Section \ref{sec:spiralproperties}, the observations of Haro 6-13 suggest that the millimeter continuum is likely optically thick. At high optical depths, the disk dust surface density cannot be well-constrained, but we can estimate a lower bound for the dust surface density and thus an upper bound for Q. 

In the absence of scattering, the dust surface density is given by

\begin{equation}
\Sigma_\mathrm{dust}(R) = \frac{\tau_\nu(R)}{\kappa_\nu^\mathrm{abs}},
\end{equation}
where $\kappa_\nu^\mathrm{abs}$ is the dust absorption coefficient. The optical depth is given by
\begin{equation}\label{eq:tau}
\tau_\nu(R) = -\cos i  \ln \left(1-\frac{I_\nu(R)}{B_\nu(T_d)} \right), 
\end{equation}
where $B_\nu(T_d)$ is the Planck function.

\begin{figure*}
\begin{center}
\includegraphics{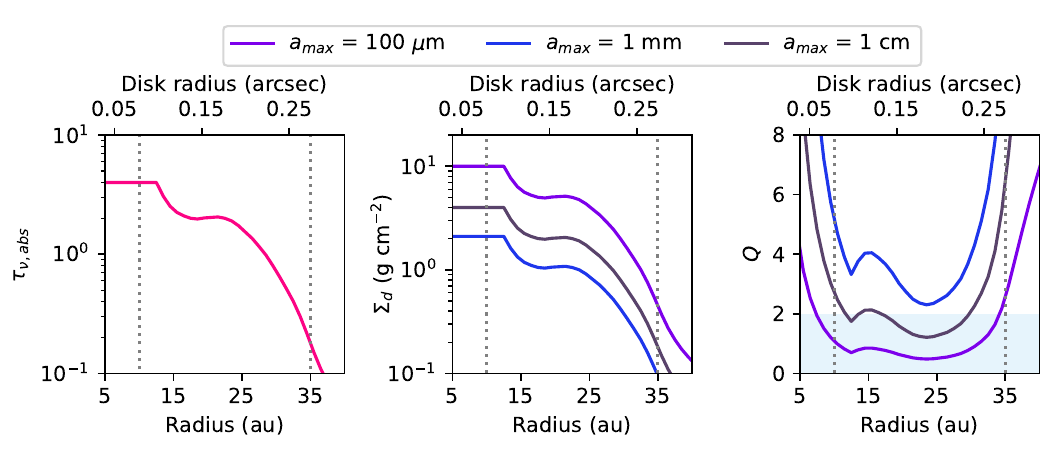}
\end{center}
\caption{Left: Absorption-only estimate for the azimuthally averaged dust optical depth at 1.3 mm. $\tau_\nu$ is fixed to 4 in the inner region where the dust temperature is set to be equal to the brightness temperature. The vertical dashed gray lines denote the approximate radial range over which the spiral arms are detected. Center: Dust surface densities calculated for various power-law grain size distributions ($n(a)\,da \propto a^{-3.5}\,da$, $a_\mathrm{max}$ = 100 $\mu$m, 1 mm, and 1 cm). Right: Corresponding Toomre $Q$ parameter estimates for the different grain size distributions. The light blue shaded region shows the range of $Q$ values for which the disk would be expected to be gravitationally unstable. \label{fig:Qplots}}
\end{figure*}

The dust temperature at each value of $R$ is calculated by taking the maximum of either the observed brightness temperature or the dust temperature calculated with Equation \ref{eq:Tdust}, since the observed brightness temperature must set the lower bound for the dust temperature. We restrict the estimate of $\tau$ to $R>5$ au because beam dilution leads to artificially low estimates of $\tau$ in the inner disk. For the regions in the inner disk where the observed brightness temperature exceeds the dust temperature calculated from Equation \ref{eq:Tdust}, we set $\tau=4$ under the assumption that $\tau$ is no lower than the values calculated outside this region. Then, $\Sigma_\mathrm{dust}$ is calculated using values of $\kappa_\nu^\mathrm{abs} = 0.4$, $1.9$, and $1.0$ cm$^{2}$ g$^{-1}$. These values correspond to the DSHARP dust opacities at a wavelength of 1.3 mm, calculated for a grain size distribution of $n(a)\,da\propto a^{-3.5}\,da$ and maximum grain sizes of $a_\mathrm{max} = 0.1$, $1$, and $10$ mm, respectively \citep{2018ApJ...869L..45B, 2018zndo...1495277B}. Finally, $Q$ is calculated assuming that $M_\ast = 0.91$ $M_\odot$ \citep{2019ApJ...882...49L} and the gas-to-dust ratio is 100. The results are plotted in Figure \ref{fig:Qplots}. Over the radial range in which the spiral arms are detected, the Toomre Q parameter values are generally low ($<4$). For certain choices of grain properties ($a_\mathrm{max}$ = 100 $\mu$m and $a_\mathrm{max}$ = 1 cm), $Q$ dips into the gravitationally unstable regime ($\lessapprox2$). In the optically thick regime, the derived value of $Q$ is highly sensitive to the temperature profile, which is not well-constrained by our data. Nevertheless, these calculations indicate that the dust optical depths do not need to be extremely high for the disk to become gravitationally unstable. In addition, multi-frequency studies of scattering in other disks have shown that absorption-only estimates of dust surface densities at millimeter wavelengths can lead to underestimates by a factor of a few \citep[e.g.,][]{2019ApJ...883...71C,2021AA...648A..33M}, so the Toomre Q values for Haro 6-13 could be significantly lower than our estimates.  Spatially-resolved observations of Haro 6-13 at multiple longer wavelengths would probe optically thinner emission and can therefore aid in inferring the dust grain properties and improve estimates of the dust surface density.

Alternatively, one might seek to estimate a disk mass from the low-resolution C$^{18}$O $2-1$ observations (the spiral arm region is not resolved in these data). The flux measured within a circle centered on the disk with a $1''$ radius (130 au) is 460 mJy km s$^{-1}$. Based on the grid of thermochemical models of compact disks presented in \citet{2021AA...651A..48M}, which assume an ISM-like CO:H$_2$ abundance, the C$^{18}$O flux translates to a maximum disk mass of $\sim10^{-2}$ $M_\odot$, corresponding to a disk-to-stellar mass ratio of $\sim0.01$. Gravitationally unstable disks are expected to have disk-to-stellar mass ratios exceeding 0.1 \citep[e.g.,][]{2016ARAA..54..271K}. However, higher resolution observations of H$_2$CO and CN exhibit negative fluxes in the inner disk, which \citet{2021AA...645A.145G} attribute to a combination of foreground absorption and high continuum optical depth. If C$^{18}$O is similarly afflicted, then it would not be a reliable probe of the surface density in the spiral arm region of the disk.

Simulations of spiral arms excited by GI have yielded different predictions for pitch angle behavior. \citet{2009MNRAS.393.1157C} and \citet{2018ApJ...860L...5F} found that the pitch angles were approximately constant, which would be at odds with our observations (under the assumption that the vertical thickness of the disk does not significantly distort the apparent geometry). However, \citet{2021ApJ...906...19C} found that for sufficiently massive and cold disks, the pitch angle could decrease with radius, similar to our observations. 

\subsubsection{Infall}
Perturbations due to infalling material can also directly excite spiral arms in a non-self-gravitating protoplanetary disk and thus influence angular momentum transport through the disk \citep{2015AA...582L...9L,2017AA...599A..86H,2025MNRAS.537.2695C}. Given the presence of an envelope around Haro 6-13, it is plausible that its spiral arms are infall-related. In the case of infall-induced spiral arms, the pitch angles are expected to increase toward the site of interaction between the disk and infalling material \citep{2025MNRAS.537.2695C}. Searching for tracers of accretion shocks, such as SO or SO$_2$ \citep[e.g.,][]{2021AA...653A.159V, 2022AA...658A.104G}, would provide insight into where the Haro 6-13 disk is being directly perturbed. 

Among the six Class I and II disks now known to have millimeter continuum spiral arms but no identified stellar or planetary companions (Elias 27, WaOph 6, IM Lup, TMC 1A, Haro 6-13, HD 143006), at least three of these (Elias 27, TMC 1A, and Haro 6-13) are known to be associated with envelope material \citep[][and this work]{2016Sci...353.1519P,2020NatAs...4..142L,2023ApJ...954..190X,2021AA...645A.145G}. HH 111 VLA 1 is an embedded disk at a projected separation of 1200 au from another star, but \citet{2020NatAs...4..142L} argue that the spiral arms of HH 111 VLA 1 are not likely to be associated with the other source due to their large separation. CO observations of IM Lup suggest that it may also have a remnant envelope \citep{2016ApJ...832..110C}. This hints that the conditions found in embedded disks and disks that are emerging from their envelopes may be facilitating the formation of millimeter continuum spiral arms. 

\subsubsection{Planetary or stellar companions}
Planetary or stellar companions can trigger spiral density waves in protoplanetary disks, with the pitch angle increasing toward the perturber \citep[e.g.,][]{1986ApJ...307..395L, 2002ApJ...565.1257T, 2018ApJ...859..119B}. However, Haro 6-13 has no known companions.  \citet{2007ApJ...662..413K} used 2MASS \citep{1990AJ.....99.1187S} to catalog wide binary systems in Taurus. No binary companions were identified for Haro 6-13 at separations between 1 and $30''$ ($\sim130-3800$ au.) Haro 6-13 is separated by $31''$ ($\sim4000$ au) from 2MASS J04321327+2429107. The former is thus sometimes referred to as Haro 6-13E in the literature, and the latter as Haro 6-13W. Nevertheless, Haro 6-13 generally has been classified as a single star because of the rarity of binaries at such wide separation \citep[e.g.,][]{2004AJ....127.1747H, 2007ApJ...662..413K, 2008AJ....135.2496C}. Even if Haro 6-13W were a bona fide companion, any spirals in the disk would be expected to be very tightly wound due to the extremely wide separation of the two stars, contrary to what is observed \citep[e.g.,][]{2018ApJ...859..118B}.   Meanwhile, high values of the Gaia renormalized unit weight error statistic (RUWE) can be checked for evidence of an unresolved companion within $1''$ \citep{2018AJ....156..259Z}. Haro 6-13's RUWE value is 2.4 \citep{2023AA...674A...1G}, within the range expected for single disk-bearing stars \citep{2022RNAAS...6...18F}. 

\citet{2020MNRAS.498.1382W} conducted a high-contrast imaging survey of disks in Taurus in the L' filter, including Haro 6-13, placing constraints on the presence of companions out to 3'' ($\sim400$ au at the distance of Haro 6-13). The survey reached a $5\sigma$ contrast of $\Delta m=6$ at $0\farcs2$ and $\Delta m\gtrapprox8$ at separations above $0\farcs5$. Assuming a hot start model and ignoring circumstellar extinction, they estimated that the median mass limit of the survey was 15 $M_J$ at 20 au and 3 $M_J$ at 150 au. Extinction from the envelope around Haro 6-13, though, may make it challenging to detect planetary or stellar mass companions from the ground. Mid-IR observations (i.e., with the James Webb Space Telescope) would be useful for searching for a companion. The Haro 6-13 disk also does not feature any clear gaps or rings that may be indicative of an embedded companion, although high optical depths at 1.3 mm may hide substructures that could be detectable in high-resolution images at longer wavelengths with the Next Generation Very Large Array (ngVLA). 

\subsubsection{Stellar flybys}
Spiral arms in disks may also result from stellar flybys \citep[e.g.,][]{1993MNRAS.261..190C, 2019MNRAS.483.4114C}. The spiral arms induced by flybys are only expected to survive for a few thousand years \citep[e.g.,][]{2019MNRAS.483.4114C}, so the perturbing star would be expected to still be nearby if the spiral arms are visible. As noted in the above discussion of stellar companions, Haro 6-13 does not have any known nearby stellar-mass perturbers. Furthermore, in systems that have been identified as likely having undergone a recent flyby, prominent extended structures that appear to be tidally induced have typically been detected in CO emission \citep[e.g.,][]{2006AA...452..897C, 2018ApJ...869L..44K, 2020ApJ...896..132Z, 2022NatAs...6..331D}. Similar structures have not been identified in molecular line observations of Haro 6-13 \citep[e.g., this work and][]{2021AA...645A.145G}. 

\subsection{Implications for compact protoplanetary disks}
Sensitive, high spatial resolution (better than $\sim10$ au) observations of protoplanetary disks have primarily targeted large disks with millimeter continuum emission extending to radii of at least 50 au and have shown that these disks are usually highly structured \citep[e.g.,][]{2018ApJ...869L..41A,2021MNRAS.501.2934C}. However, the majority of millimeter continuum disks, including Haro 6-13, appear to be more compact \citep[e.g.,][]{2019ApJ...882...49L, 2020ApJ...895..126H}. The faint, extended emission observed in the Haro 6-13 disk (Figure \ref{fig:radialprofile}) may bring into question how disk sizes should be defined (conventionally based on the radius enclosing some percentage of the flux), but this could also be an issue for many other disks that have been described as ``compact'' because most have only been observed at moderate sensitivity. Several recent studies have targeted compact disks with ALMA's long baselines in order to examine to what extent the kinds of substructures commonly detected in large disks are also present in small disks, generally turning up a mixture of ringed and apparently featureless disks \citep[e.g.,][]{2019AA...626L...2F,2024AA...682A..55M,2025AA...696A.232G}.

If the Haro 6-13 disk is determined to be gravitationally unstable, it would be the smallest such disk identified to date. The other Class I and II disks for which gravitational instability has been suggested as a possible origin for their millimeter continuum spiral arms (IM Lup, Wa Oph 6, Elias 27, TMC 1A, HH 111 VLA 1, MWC 758) have had disk emission detected at large radii, beyond 100 au \citep{2016Sci...353.1519P, 2018ApJ...860..124D,2018ApJ...869L..43H,2020NatAs...4..142L, 2023ApJ...954..190X}. Alternatively, if Haro 6-13's spiral arms result from stellar flybys or direct perturbations from infall in the absence of GI, their radial extent would suggest that external perturbations can affect disk dynamics in the midplane on scales comparable to the semi-major axes of the Solar System giant planets.  Millimeter continuum spiral arms have seldom been resolved inside compact disks (although some compact disks are known to have spiral arms surrounding them in extended circumbinary material \citep[e.g.,][]{2014ApJ...796....1T, 2020ApJ...898...10T, 2022ApJ...930...91D}). The notable exceptions are the HT Lup A, AS 205 N, and HD 100453 disks, which have millimeter continuum radii of $\sim30-60$ au and stellar companions that are orbiting outside the spiral arms at projected separations ranging from $\sim25-170$ au \citep{ 2018ApJ...869L..44K, 2020MNRAS.491.1335R}. The small sizes of these disks may be due to truncation by the companions \citep[e.g.,][]{1993MNRAS.261..190C,1994ApJ...421..651A}. After DSHARP, which detected two of the known cases of millimeter continuum spiral arms inside compact disks, high resolution millimeter continuum disk surveys have generally been at lower sensitivity \citep[e.g.,][]{2021MNRAS.501.2934C, 2024ApJ...966...59S, 2025AA...696A.232G}. While gaps and rings can be high-contrast enough to be detectable at relatively low sensitivities, millimeter continuum spiral arms typically have low contrasts \citep[e.g.,][]{2018ApJ...869L..43H}. The detection of spiral arms in our new observations of the Haro 6-13 disk, which is not known to have a companion, motivates further deep, high resolution imaging of the compact disk population without known companions to ensure that searches for spiral arms are appropriately inclusive and to improve our understanding of the kinds of systems in which they are prevalent. 

High-resolution studies of several disks have found evidence that they are optically thick in their inner 20-30 au at millimeter wavelengths \citep[e.g.,][]{2021AA...648A..33M, 2023AA...673A..77R,2024NatAs...8.1148U}. Haro 6-13 likewise appears to be optically thick interior to 30 au. It has been questioned whether there is a discrepancy between protoplanetary disk masses and exoplanet masses \citep{2014MNRAS.445.3315N, 2018AA...618L...3M}. However, high optical depths can lead to significant underestimates of disk masses and greater difficulties in detecting substructures. This problem points to a use case for the ngVLA, which would be able to obtain high resolution images of disks at longer wavelengths, where the optical depth is lower and the dust surface density profiles can therefore be measured more accurately. However, free-free contamination may pose a challenge to characterizing dust properties at these longer wavelengths.

\subsection{Haro 6-13 as a (mild) cautionary tale}
Given the lengthy integration times and good weather required to obtain high signal-to-noise long-baseline ALMA observations, it has been popular to explore how to maximize the information that can be extracted from lower-resolution observations. Visibility modeling and super-resolution imaging techniques have proven to be successful in discerning the presence of disk substructures that are not readily apparent in the CLEAN images \citep[e.g.,][]{2016ApJ...818L..16Z, 2020MNRAS.495.3209J,2023PASP..135f4503Z}. 

Our new Haro 6-13 observations, though, also highlight some of the limitations of the aforementioned techniques. Searches for substructures via visibility modeling have mostly assumed that the disk is axisymmetric \citep[e.g.,][]{2016ApJ...818L..16Z, 2020MNRAS.495.3209J, 2023ApJ...952..108Z}. High-resolution disk images suggest that this assumption is reasonable most of the time \citep[e.g.,][]{2018ApJ...869L..42H,2021MNRAS.501.2934C}. However, when azimuthally averaged, disks with spiral arms may appear as though they have (additional) gaps and rings \citep{2022MNRAS.509.2780J}. Consequently, the presence of protoplanets may be erroneously inferred if one is not aware of the spiral arms. In addition, Haro 6-13's spiral structures appear ring-like in super-resolution images made from observations with baselines up to $\sim3700$ km and $\sim4$ minutes on source \citep{2024PASJ...76..437Y}. Hence, our sensitive ALMA observations with $4\times$ longer baselines have been essential for understanding the true morphology of the Haro 6-13 disk. As noted by \citet{2018ApJ...869L..41A}, a spatial resolution of 5 au or better is desirable for investigating disk substructures because it is comparable to the expected pressure scale height at several tens of au. Models indicate that the scale height is closely linked to the characteristic size scales of substructures \citep[e.g.,][]{1999ApJ...514..344B, 2009ApJ...697.1269J}.  Nevertheless, even if visibility modeling and super-resolution imaging do not capture the exact nature of the substructures present, they can still be useful for identifying systems that would be interesting for follow-up at higher resolution and sensitivity. 

\section{Summary} \label{sec:summary}
We present the highest resolution millimeter continuum observations to date of the Haro 6-13 protoplanetary disk. A pair of low-contrast spiral arms is detected, making the Haro 6-13 disk one of the smallest disks in which millimeter continuum spiral arms have been observed. The arms can be modelled as Archimedean spirals with pitch angles ranging from $\sim10-30^\circ$. The low value of the $229.5-97.5$ GHz (1.3-3 mm) spectral index ($\alpha=2.1$) and the high brightness temperatures provide evidence that the millimeter continuum is optically thick. The evidence for high optical depths and the presence of envelope material motivate further exploration of GI as an explanation for the spiral arms. In the absence of GI, perturbations from infall may still contribute to spiral arm formation. The Haro 6-13 disk helps to demonstrate that diverse substructures can be found in compact protoplanetary disks and points to the need for further observations of spatially resolved, optically thin disk tracers in order to probe the mass reservoir in the inner tens of au. 

\section*{Acknowledgements}
 This paper makes use of the following ALMA data: ADS/JAO.ALMA\#2022.1.01365.S and \#2016.1.01042.S. ALMA is a partnership of ESO (representing its member states), NSF (USA) and NINS (Japan), together with NRC (Canada), MOST and ASIAA (Taiwan), and KASI (Republic of Korea), in cooperation with the Republic of Chile. The Joint ALMA Observatory is operated by ESO, AUI/NRAO and NAOJ.  We also thank our contact scientist Ryan Loomis for his assistance. MB has received funding from the European Research Council (ERC) under the European Union's Horizon 2020 research and innovation programme (PROTOPLANETS, grant agreement No. 101002188). SF is funded by the European Union (ERC, UNVEIL, 101076613), and acknowledges financial contribution from PRIN-MUR 2022YP5ACE. Views and opinions expressed, however, are those of the author(s) only and do not necessarily reflect those of the European Union or the ERC. Neither the European Union nor the granting authority can be held responsible for them. The research of MK is funded by a Reintegration fellowship of the Carlsberg Foundation (CF22-1014).

\facilities{ALMA}

\software{\texttt{analysisUtils} \citep{2023zndo...7502160H}, 
\texttt{AstroPy} \citep{2013AA...558A..33A, 2018AJ....156..123A, 2022ApJ...935..167A}, 
\texttt{CASA} \citep{2022PASP..134k4501C}, 
\texttt{cmasher} \citep{cmasher}, 
\texttt{dsharp\_opac} \citep{2018zndo...1495277B},
\texttt{emcee} \citep{2013PASP..125..306F},
\texttt{filfinder} \citep{2015MNRAS.452.3435K, 2016ascl.soft08009K}, 
\texttt{george} \citep{2015ITPAM..38..252A},
 \texttt{matplotlib} \citep{Hunter:2007},  
 \texttt{protomidpy} (\citealt{2024MNRAS.532.1361A}, \url{https://github.com/2ndmk2/protomidpy}),
  \texttt{scikit-image} \citep{scikit-image},
 \texttt{visread} \citep{ian_czekala_2021_4432520},
 \texttt{SciPy} \citep{2020SciPy-NMeth}}

\end{CJK*}
\end{document}